\documentclass[aps,twocolumn,amssymb,prl,showpacs,10pt]{revtex4-1}
\usepackage{graphicx}
\usepackage{multirow}
\usepackage[breaklinks,colorlinks,urlcolor=blue,citecolor=blue]{hyperref}
\usepackage{mathrsfs}
\usepackage{amsmath}
\usepackage{soul}
\usepackage{color}
\usepackage[normalem]{ulem}
\usepackage{float}
\usepackage[utf8]{inputenc}
\usepackage{lmodern}

\begin{document}

\title{Combining Electromagnetic and Gravitational-Wave Constraints
on Neutron-Star Masses and Radii}
\author{Mohammad Al-Mamun$^{1}$}
\author{Andrew W. Steiner$^{1,2}$} 
\author{Joonas N\"{a}ttil\"{a}$^{3,4}$} 
\author{Jacob Lange$^{5}$}
\author{Richard O'Shaughnessy$^{5}$}
\author{Ingo Tews$^{6}$}
\author{Stefano Gandolfi$^{6}$}
\author{Craig Heinke$^{7}$}
\author{Sophia Han$^{8,9}$}
\affiliation{$^{1}$Department of Physics and Astronomy, University of
  Tennessee, Knoxville, TN 37996, USA}
\affiliation{$^{2}$Physics Division, Oak Ridge National Laboratory, Oak
  Ridge, TN 37831, USA}
\affiliation{$^{3}$Physics Department and Columbia Astrophysics
  Laboratory, Columbia University, 538 West 120th Street,
  New York, NY 10027, USA}
\affiliation{$^{4}$Center for Computational Astrophysics,
  Flatiron Institute, 162 Fifth Avenue, New York, NY 10010, USA}
\affiliation{$^{5}$ Rochester Institute of Technology, 85 Lomb Memorial Drive,
Rochester, NY 14623, USA}
\affiliation{$^{6}$ Theoretical Division, Los Alamos National Laboratory, Los Alamos, NM 87545, USA}
\affiliation{$^{7}$ Department of Physics, CCIS 4-183, University of Alberta, Edmonton, AB, T6G 2E1, Canada}
\affiliation{$^{8}$ Department of Physics, University of California, Berkeley, CA 94720, USA}
\affiliation{$^{9}$ Department of Physics and Astronomy, Ohio University, Athens, OH 45701, USA}

\begin{abstract}
  We perform a joint Bayesian inference of neutron-star mass and
  radius constraints based on GW170817, observations of quiescent
  low-mass X-ray binaries (QLMXBs), photospheric radius expansion
  X-ray bursts (PREs), and X-ray timing observations of J0030+0451.
  With this data set, the form of the prior distribution still has an
  impact on the posterior mass-radius (MR) curves and equation of
  state (EOS), but this impact is smaller than recently obtained when
  considering QLMXBs alone. We analyze the consistency of the
  electromagnetic data by including an ``intrinsic scattering''
  contribution to the uncertainties, and find only a slight broadening
  of the posteriors. This suggests that the gravitational-wave and
  electromagnetic observations of neutron-star structure are providing
  a consistent picture of the neutron-star mass-radius curve and the
  EOS.
\end{abstract}
\pacs{97.60.Jd, 95.30.Cq, 26.60.-c}
\maketitle


The idea that neutron stars (NSs) might be useful in determining the
equation of state (EOS) of dense matter~\cite{Cameron59ns} precedes
the discovery of the first NS in 1967~\cite{Hewish68oo} by almost a
decade. Until recently, the strongest observational constraints on the
EOS came from NS mass measurements. These mass measurements all lay in
a narrow range around $1.4~\mathrm{M}_{\odot}$~\cite{Thorsett99ns}
until the last decade, when NSs with masses near
$2.0~\mathrm{M}_{\odot}$ were discovered~\cite{Barziv01tm, Demorest10,
  vanKerkwijk11ef, Antoniadis13, Cromartie20rs}. Measurements of NS
radii, on the other hand, have been plagued with various systematic
uncertainties~\cite{Lattimer07}. The past decade has seen an
increasing number of observations which constrain both the NS mass and
radius with better-controlled systematic uncertainties, providing
stronger EOS constraints. Quiescent low-mass X-ray
binaries~\cite{Heinke06} (``QLMXBs''), NSs which exhibit photospheric
radius expansion X-ray bursts~\cite{Ozel09} (``PREs''), and nearby
isolated NSs~\cite{Pons02,Ho09} all have been used to provide mass and
radius measurements (see e.g. Refs.~\cite{Lattimer12,Ozel16} for
recent reviews). Importantly, these measurements rely on data from
various different instruments and connect the radius to the actual
observables using different theoretical assumptions. These in turn all
yield different underlying systematics. Finally, the recent
observations of gravitational waves (GW) from binary NS mergers
(GW170817 and GW190425)~\cite{Abbott17go, Abbott18mo, Abbott:2020uma}
by the LIGO Scientific- and Virgo collaborations (LVC) or X-ray
observations of J0030+0451 by the NICER
collaboration~\cite{Riley2019,Miller19} provide additional information
on the EOS.

There are several recent works which analyze the data from GW170817
and/or GW190425 as well as NICER and its implication on the NS
EOS~\cite{Annala18, Fattoyev18, BKumar19, Coughlin:2018miv,
  Coughlin:2018fis, Tews:2018iwm, Raithel19co, De18, Radice2019,
  Capano20, Baillot2019, Jiang19, Dietrich:2020lps, Landry19,
  Landry20, Essick:2020flb, Jiang20, Lattimer2019, Ayrian18,
  Raaijmakers:2019dks}, but very few (e.g.,
Ref.~\cite{Baillot2019,Raithel20os}) directly combine the GW data with
constraints on NS radii from QLMXBs and PREs. However, these
additional observational sources add valuable information.
Ref.~\cite{Steiner15un} used QLMXBs and PREs observations to
{\em{predict}} the NS tidal deformability which would be inferred from
GW observations. They predicted the dimensionless tidal deformability
of a $1.4~\mathrm{M}_{\odot}$ NS, $\Lambda_{M=1.4}$, was between $130$
and $460$ to 95\% confidence. The recent analysis of GW
170817~\cite{Abbott18mo} by the LVC found
$\Lambda_{M=1.4}\in[70,580]$, to 90\% confidence, matching the
prediction to within errors.

In this letter, we present a Bayesian inference of the
NS structure data, including both GW data as well as data from
electromagnetic observation of QLMXBs and PREs, using less restrictive
assumptions than made in previous works.


We build upon the method first described in
Ref.~\cite{Steiner10te,Steiner12cn} (see also Ref.~\cite{Read09co}),
reviewed in Ref.~\cite{Lattimer14co}, and detailed in the Supplemental
Material. Ref.~\cite{Steiner13tn} first demonstrated that the choice
of EOS parameterization has a significant impact on both the posterior
mass-radius relation and the EOS, see also Ref.~\cite{Greif:2018njt}.
To estimate the impact of that choice, here as in
Ref.~\cite{Steiner18ct}, our prior distribution is built on two EOS
parameterizations: (a) one which uses three polytropes (referred to as
``3P'') and (b) one which uses four line segments in the space of
pressure vs. energy density (referred to as ``4L''). The latter
parametrization has a stronger preference for strong phase transitions
(regions where the pressure is nearly independent of the energy
density). We use these parameterizations because both of them are
physically reasonable, yet they give qualitatively different
mass-radius curves. This allows us to study how the prior choice
creates additional uncertainty in our results.

At each point in our EOS parameter spaces, we solve the
Tolman-Oppenheimer-Volkov (TOV)
equations~\cite{Tolman:1939jz,Oppenheimer:1939ne}, compute the moment
of inertia as a function of the central pressure, and use the
Yagi-Yunes (YY) correlation~\cite{Yagi13} (as formulated in
Ref.~\cite{Steiner16ns}) to compute the tidal deformability as a
function of central pressure. This method of computing the tidal
deformability is much faster than a direct computation, and while
deviations from the exact result of up to 10\% are
possible~\cite{Han19td,Carson:2019rjx}, the correlation is accurate to
within a few percent for the typical EOS in our prior and posterior
distributions. We construct the conditional probability for QLMXBs as
in Ref.~\cite{Steiner18ct} and generalize it to include 3 PRE X-ray
bursting sources~\cite{Nattila16eo,Nattila17ns}. We also include the
LIGO constraints on $\tilde{\Lambda}$ from GW170817, using the
SEOBNRv4T model for binary NS inspiral assuming low compact object
spins~\cite{Abbott19}. GW observations are incorporated by tabulating
and interpolating a marginal likelihood versus the masses for each
object and the combined tidal deformability. Marginalization is
performed via RIFT~\cite{RIFT} over all extrinsic parameters and our
fiducial low-spin prior assumptions. Finally, we also include mass and
radius constraints on J0030+0451 from the NICER
instrument~\cite{Riley2019,Miller19}. We reject all EOSs which are
acausal or imply a maximum NS mass less than $2~\mathrm{M}_{\odot}$.
The mass cutoff does have a small uncertainty, but this uncertainty
does not qualitatively impact our results. Also, we could have chosen
to replace one of the high-density EOS parameters with
$M_{\mathrm{max}}$, but since neither the high-density part of the EOS
nor the maximum mass is well-known, this prior choice is not
necessarily better (or worse) than ours.

While it is often helpful to directly model systematic uncertainties,
as was partially done in earlier
works~\cite{Nattila16eo,Nattila17ns,Steiner18ct}, these estimations
require detailed models for the uncertainties which may not be
perfect. Systematic uncertainties may result in intrinsic scattering
(IS) which we model by convolving the probability distribution ${\cal
  D}(R,M)$ for each star with a Gaussian kernel. This addition of an
extra uncertainty to the observations allows us to {\em{quantify}} the
possible level of systematic bias present in each measurement: the
value of each IS parameter is expected to increase until the full
dataset used in the inference is self-consistent. We emphasize that
there is no reason to believe that a Gaussian is necessarily the
{\em{correct}} distribution for additional unknown systematics, but we
believe it is a reasonable first guess. Our approach is based on the
intrinsic scattering parameter from Ref.~\cite{Nattila17ns} which was
applied to the X-ray spectrum rather than the mass and radius
constraints. The details of this procedure and the Markov Chain Monte
Carlo (MCMC) method are given in the Supplemental Material.

We argue that neither of the two EOS parameterizations are more or
less motivated by quantum chromodynamics (QCD), and thus we assign
them equal prior probability. We present the posteriors for these two
EOS parameterizations, 3P and 4L, separately to make the impact of a
different EOS parameterization clear. We also present four different
data sets: (i) GW170817 only (GW), (ii) GW170817$+$QLMXBs$+$PREs (GW,
QLMXB, PRE), (iii) GW170817$+$QLMXBs$+$PREs$+$NICER (all), and (iv)
GW170817$+$QLMXBs$+$PREs+NICER with an additional intrinsic scattering
parameter added to each of the QLMXBs, PREs and NICER (all+IS). We
thus performed 8 different sets of simulations.

\begin{figure}
  \includegraphics[width=0.99\linewidth]{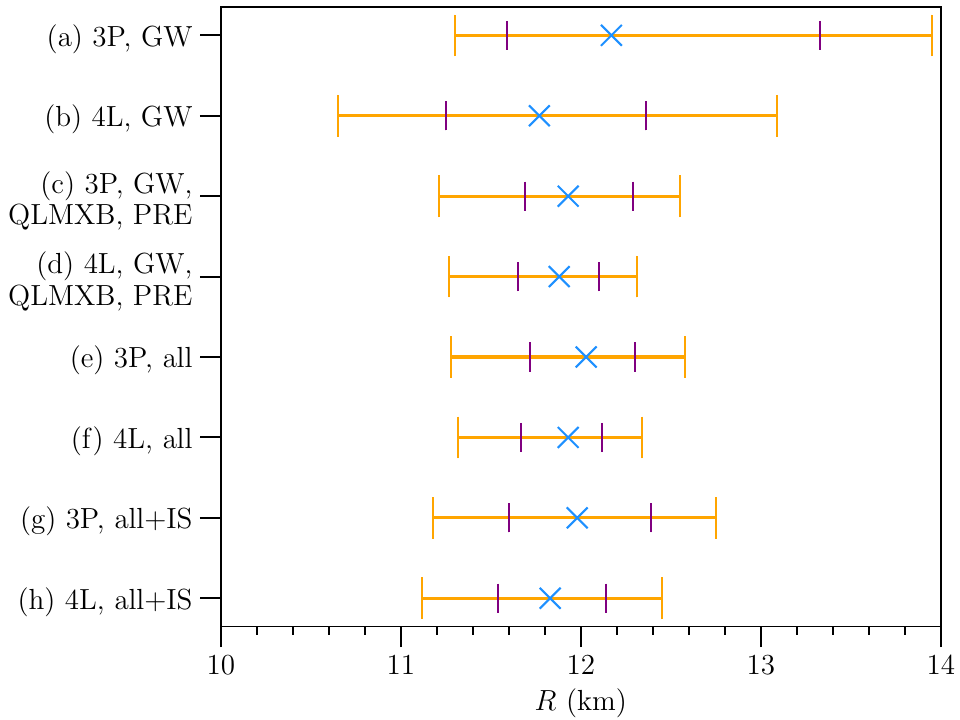}
  \caption{{\label{fig:model_radii}} Radius measured for a
    $1.4~\mathrm{M}_{\odot}$ NS in different models. Blue crosses
    indicate the median points, while the purple and orange bars
    represent the $68\%$ and $95\%$ credible intervals.}
\end{figure}

Posteriors for the radius of a 1.4 solar mass NS, $R_{1.4}$, are
summarized in Figure~\ref{fig:model_radii}. As might be expected, the
limits for the GW data alone [(a) and (b)] are the least
constraining. The 95\% credible intervals are 11.3$-$13.9 km (3P) and
10.7$-$13.1 km (4L). Choosing a line segment-based EOS prior decreases
the lower (upper) radius limit by 0.5 (1) km. The next four bar plots
in the figure [(c) - (f)] show that the range shrinks significantly when adding 
the EM observations without IS. In addition, once the EM observations
are added, these constraints are less sensitive to the EOS prior than
the QLMXB observations alone~\cite{Steiner18ct}. Including the IS
contribution [(g) and (h)] slightly broadens
the constraints coming from the EM data.

\begin{figure}
  \includegraphics[width=0.99\linewidth]{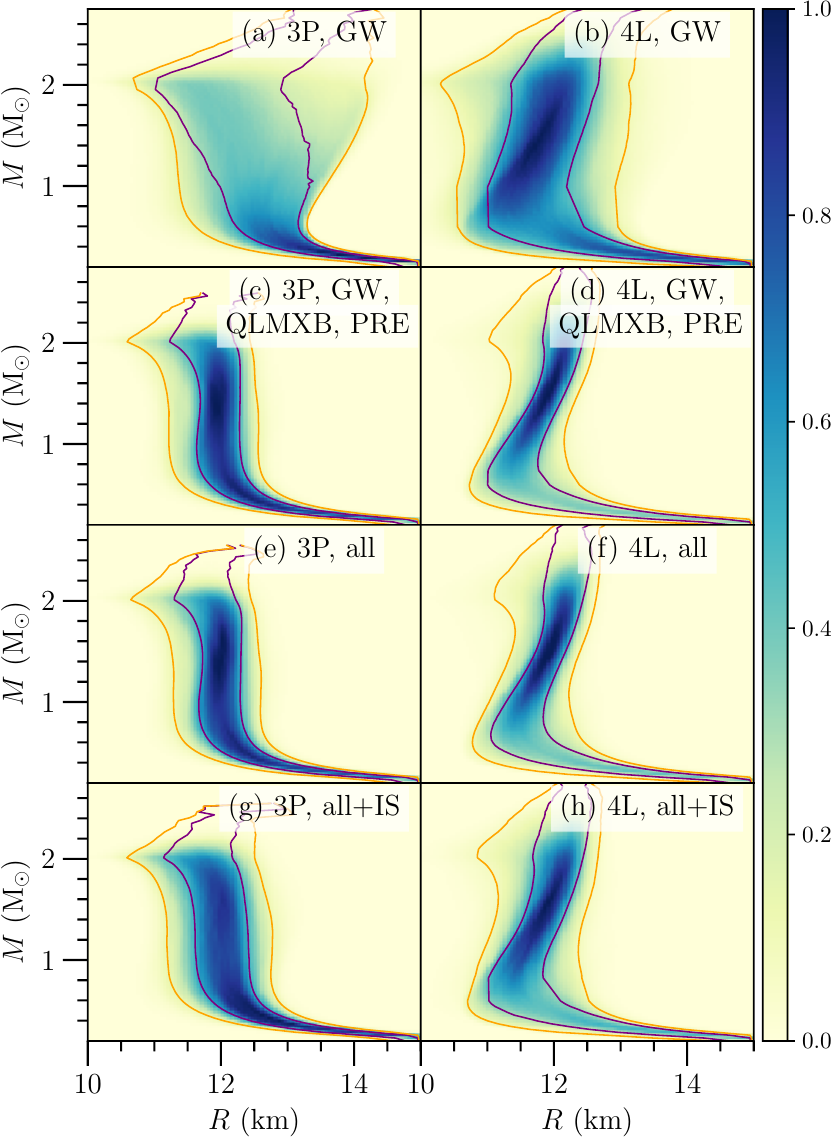}
  \caption{\label{fig:mvsr}Posterior distributions for the NS radius
    as a function of the gravitational mass. Left panels are
    constructed with the ``3P'' EOS and right panels with the ``4L''
    EOS. Different rows refer to different data selections.}
\end{figure}

Figure~\ref{fig:mvsr} shows the posterior distributions for the NS
radius as a function of the gravitational mass. The shape of the M-R
curve is more sensitive to the EOS prior than the radius of a 1.4
solar mass NS alone. The 4L EOS prior, because of the potential for
phase transitions to modify the EOS at low densities, produces smaller
radii for low-mass stars and larger radii for high-mass stars. This
distinction may be particularly important in light of a possible 2.6
$\mathrm{M}_{\odot}$ NS in GW190814~\cite{Abbott:2020khf,
  Essick:2020ghc, Tews:2020ylw}.

\begin{figure}
  \includegraphics[width=0.99\linewidth]{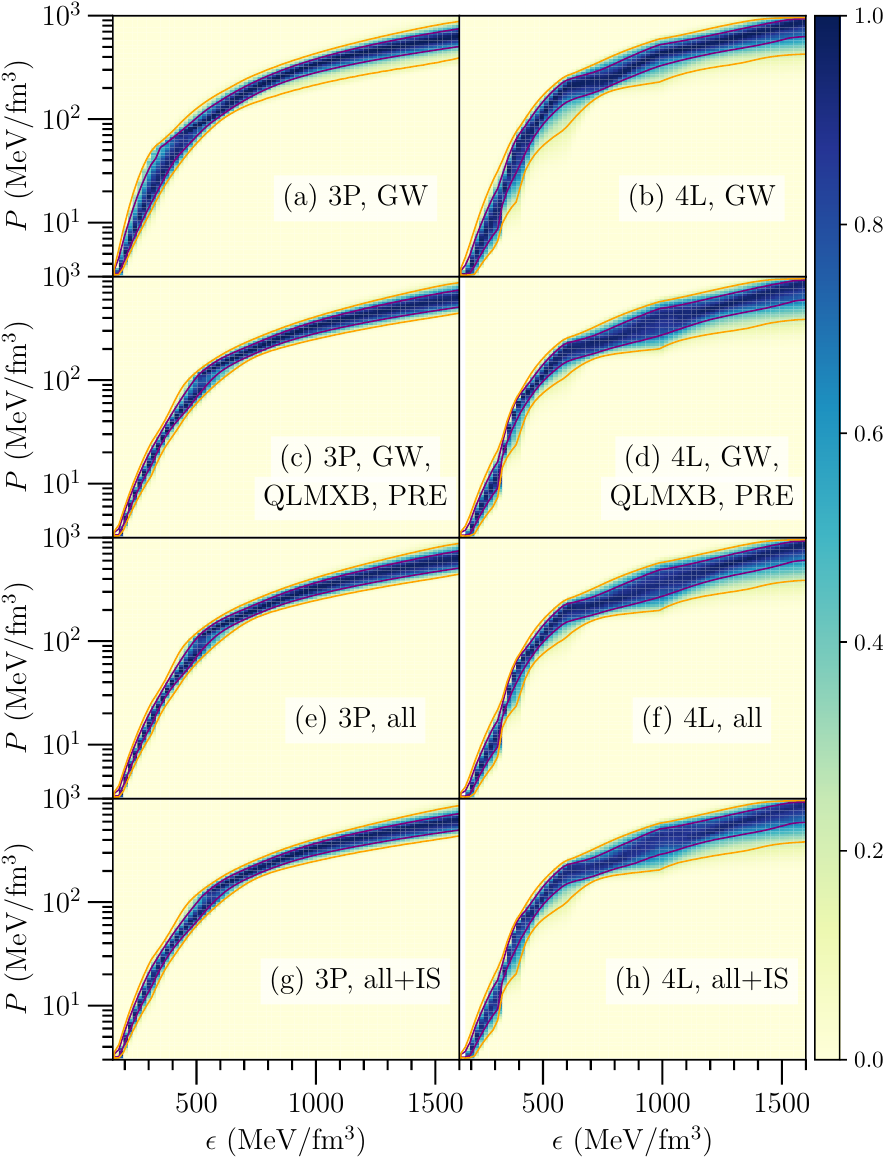}
  \caption{\label{fig:eos}Posterior distributions for the
    pressure as a function of the energy density. Left panels are
    constructed with the ``3P'' EOS and right panels with the ``4L''
    EOS. Different rows refer to different data selections.}
  \label{fig:pvse}
\end{figure}

\begin{figure}
  \includegraphics[width=0.99\linewidth]{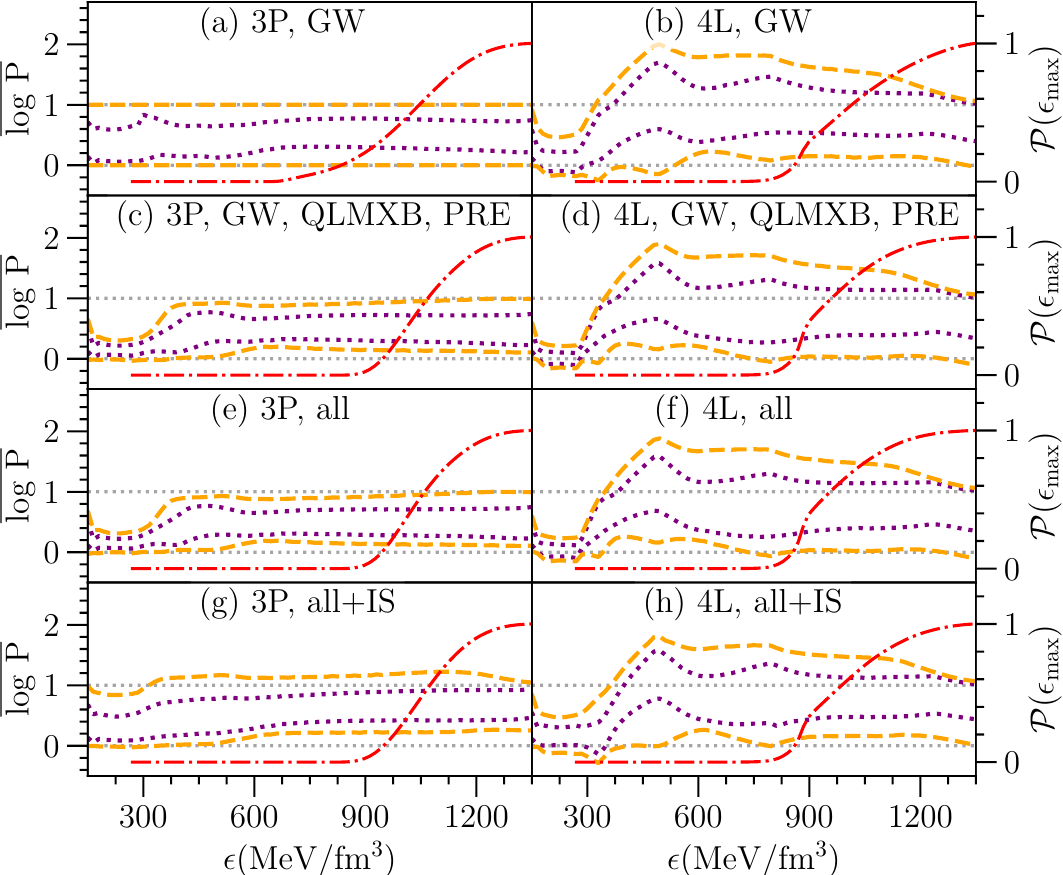}
  \caption{{\label{fig:eos_normed}} The 68\% (purple dotted) and 95\%
    (orange dashed) credible intervals for the pressure as a function
    of the energy density. All intervals are modified with the same
    linear transformation to ensure that the upper and lower 95\%
    intervals in panel (a) are always at 0 and 1. The right-hand y
    axis~(red dot-dashed line) shows the posterior probability that
    the central energy density is smaller than the value on the x
    axis.}
\end{figure}

Figure~\ref{fig:pvse} shows the posteriors for the pressure as a
function of the energy density. The combination of the EM and GW data
strongly constrains the pressure until about 400-500
$\mathrm{MeV}/\mathrm{fm}^{3}$. In order to more easily compare the
pressures between different models or data sets,
Figure~\ref{fig:eos_normed} shows an alternate version where, for each
energy density, all 8 panels are rescaled and shifted by the same
linear transformation which ensures that the 95\% credible intervals
for panel (a) lie exactly at 0 and 1. Thus, for panels (b) through
(h), the orange dashed curves show the change in pressure of different
models relative to that in panel (a). For reference, the energy
density at nuclear saturation is $\epsilon_0\approx
150~\mathrm{MeV}/\mathrm{fm}^{3}$. The red dot-dashed curves show the
probability that the central energy density of the maximum mass NS is
smaller than the energy density from the x-axis. Most EOSs suggested
by the data imply the central energy density is between about 900 and
$1200~\mathrm{MeV}/\mathrm{fm}^{3}$. In the 3P model, the EM data
suggests a smaller pressure for small energy densities and almost
unchanged at higher densities. In the 4L model, the effect is more
dramatic: a smaller pressure at low densities is compensated for by an
increase in the pressure at higher densities. For
$\epsilon>1200~\mathrm{MeV}/\mathrm{fm}^{3}$, it is unlikely the data
is strongly constraining the EOS, i.e. the EOS is strongly impacted by
the prior distribution.

We compare our results to several other previous works in
Table~\ref{tab:rcomp}. Our results on $R_{1.4}$ with the GW data alone
are consistent with previous works which include only limited
information from NS radius constraints~\cite{Annala18, Ayrian18, De18,
  Radice2019, Raithel19co, BKumar19, Capano20}. The variation in these
results across the various references is consistent with our finding
that these posterior distributions depend on the EOS prior
distribution (as well as on the other details of the analysis).

Other works find, as we do, that the radius constraints are tighter
when the EM data is included. Our results which include IS suggest
(but do not definitively prove) that this result is not due to
systematic uncertainties which are artificially constraining
NS radii.

Gravitational wave observations have suggested other possible more
indirect constraints on the EOS, and we summarize the impact of some
of these constraints have on the radius of a $1.4~\mathrm{M}_{\odot}$
NS in Table~\ref{tab:cnst_mr} by applying them to our ``all+IS''
posteriors. Refs.~\cite{Radice17gj,Radice2019} found that GW170817
implied a lower limit on $\tilde{\Lambda}>300$, because EOSs with
smaller values of $\tilde{\Lambda}$ do not create a sufficiently
massive accretion disk to the observed kilonova. This constraint may
be optimistic (e.g. see Ref.~\cite{Bauswein17sd}) (and see also a
different perspective in Ref.~\cite{Kiuchi19rt}), but it has a
relatively low impact. It increases the lower limit on the 95\%
confidence limit for $R_{1.4}$ by about 0.2 km. Ref.~\cite{Ruiz18}
found that $M_{\mathrm{max}} \leq 2.17~\mathrm{M}_{\odot}$ was
required to ensure that no short-lived hypermassive NS (which was not
observed) was present in GW170817. We find that this constraint has
only a weak impact on our posterior distributions. The maximum mass is
weakly correlated by the radius of a 1.4 solar mass NS, as shown the
Supplemental Material, so when we decrease the typical maximum mass we
also slightly decrease the typical radius. Finally, GW190814 implied
the merger of a $2.6~\mathrm{M}_{\odot}$ object with a more massive
black hole. Our MCMC simulation for the 3P model generated no
configurations with maximum masses this large. This does not
necessarily mean that 3P parameter sets with large maximum masses do
not exist, but they do appear highly improbable. In the 4L model,
increasing $M_{\mathrm{max}}$ to $2.6~\mathrm{M}_{\odot}$ increases
the 95\% lower limit for the radius by 0.5 km. For example, in
Table~\ref{tab:cnst_mr}, changing the $M_{\mathrm{max}}$ value in 4L
models increased $R$ lower limit of $95\%$ C.I. from 10.98 km to 11.47
km for a 1.4 $M_{\odot}$ NS.
  
\begin{table}
  \begin{tabular}{c|ccl}
    \hline
    Reference & $\mathrm{R_{1.4}}$ & C.I. & Source \\
    \hline
    \cite{Abbott18mo} & [10.5, 13.3] & $90\%$ & GW \\
    \cite{Annala18} & [9.9, 13.6] & $90\%$ & GW \\
    \cite{Fattoyev18} &  $<13.6$ & $90\%$ & GW \\
    \cite{BKumar19} & [9.4, 12.8] & $90\%$ & GW \\
    \cite{Raithel19co} & [9.8, 13.2]\footnote{Radius measurement
      for the primary NS of the merger event} & $90\%$ & GW \\
    \cite{Essick:2020flb} & [10.36, 12.78] & $90\%$ & GW \\
    Model ``a'' & [11.30, 13.95] & $95\%$ & GW \\
    Model ``b'' & [10.65, 13.09] & $95\%$ & GW \\
    \cite{De18} & [8.9, 13.2] & $90\%$ & GW, merger remnant \\
    \cite{Radice2019} & [11.4, 13.2] & $90\%$ & GW, merger remnant \\
    \cite{Capano20} & [10.4, 11.9] & $90\%$ & GW, merger remnant \\
    \citep{Baillot2019} & [11.98, 12.76] & $90\%$ & GW, QLMXB \\
    \citep{Jiang19} & [10.5, 11.8] & $90\%$ & GW, QLMXB \\
    \cite{Dietrich:2020lps} & [10.94, 12.72] & $90\%$ &
    GWs\footnote{GWs referred to the joint analysis of GW170817
      and GW190425}, NICER \\
    \cite{Landry19,Landry20} & [10.85, 13.41] & $90\%$ & GWs, NICER \\
    \cite{Essick:2020flb} & [11.91, 13.25] & $90\%$ & GW, NICER \\
    \cite{Jiang20} & [11.3, 13.3] & $90\%$ & GW, NICER \\
    \cite{Raithel20os} & [12, 13] & $90\%$ & GWs, NICER \\
    \cite{Raithel20os} & [10.0, 11.5] & $90\%$ & GWs, QLMXB, PRE \\
    Model ``c'' & [11.21, 12.55] & $95\%$ & GW, QLMXB, PRE \\
    Model ``e'' & [11.28, 12.58] & $95\%$ & GW, QLMXB, PRE, NICER \\
    \hline
  \end{tabular}	
  \caption{\label{tab:rcomp} A comparison of our posterior
    distributions for the radius of a $1.4~\mathrm{M}_{\odot}$ NS in
    comparison to other results obtained in the literature.}
\end{table}

\begin{table}
  \begin{center}
    \begin{tabular}{c | c c c c c }
      \hline
      Model \& constraints & $-2\sigma$ & $-1\sigma$ &
      med. & $+1\sigma$ & $+2\sigma$ \\
      \hline
      3P, all+IS & 11.18 & 11.6 & 11.98 & 12.39 & 12.75 \\
      4L, all+IS & 11.12 & 11.54 & 11.83 & 12.14 & 12.45 \\
      3P, all+IS~($\tilde{\Lambda} > 300$) &
      11.33 & 11.63 & 11.98 & 12.37 & 12.71 \\
      4L, all+IS~($\tilde{\Lambda} > 300$) &
      11.28 & 11.56 & 11.83 & 12.12 & 12.40 \\
      3P, all+IS~($M_{\mathrm{max}} < 2.17~\mathrm{M}_{\odot}$) &
      11.15 & 11.54 & 11.98 & 12.31  & 12.72 \\
      4L, all+IS~($M_{\mathrm{max}} < 2.17~\mathrm{M}_{\odot}$) &
      10.98 & 11.43 & 11.88 &12.13 & 12.46 \\
      4L, all+IS~($M_{\mathrm{max}} > 2.6~\mathrm{M}_{\odot}$) &
      11.47 & 11.76 & 11.98 & 12.19 & 12.42 \\	
      \hline
    \end{tabular}	
    \caption{\label{tab:cnst_mr} The $1\sigma$ and $2\sigma$
      confidence limits with the median for the radius of
      $1.4~\mathrm{M}_{\odot}$ NS in km with applied constraints on
      intrinsic scattering models. }
  \end{center}
\end{table}

\begin{acknowledgments}
  M.A. was supported by NSF grant AST 1909490. A.W.S. was supported by
  NSF grant PHY 1554876, by the U.S. DOE Office of Nuclear Physics,
  Nordic Institute for Theoretical Physics (NORDITA), and the
  University of Turku. J.L. and R.O.S. were supported by NSF grants
  PHY 1707965 and AST 1909534. I.T. and S.G. were supported by the
  U.S. Department of Energy, Office of Science, Office of Nuclear
  Physics, under contract No.~DE-AC52-06NA25396, by the NUCLEI SciDAC
  program, and by the LDRD program at LANL. S.G. was also supported by
  the DOE Early Career research Program. C.H. is supported by NSERC
  Discovery Grant RGPIN-2016-04602, and a Discovery Accelerator
  Supplement. S.H. is supported by the National Science Foundation,
  Grant PHY-1630782, and the Heising-Simons Foundation, Grant
  2017-228. This work used the Extreme Science and Engineering
  Discovery Environment (XSEDE) allocation PHY 170048 and PHY 180052
  supported by NSF grant number ACI-1548562. The open-source code for
  this work~\cite{Steiner14ba} is built upon
  O$_2$scl~\cite{Steiner14oo}, GSL, HDF5, FFTW~\cite{Frigo12}, and
  matplotlib.
\end{acknowledgments} 

\bibliographystyle{apsrev}
\bibliography{paper}

\begin{thebibliography}{69}
\expandafter\ifx\csname natexlab\endcsname\relax\def\natexlab#1{#1}\fi
\expandafter\ifx\csname bibnamefont\endcsname\relax
  \def\bibnamefont#1{#1}\fi
\expandafter\ifx\csname bibfnamefont\endcsname\relax
  \def\bibfnamefont#1{#1}\fi
\expandafter\ifx\csname citenamefont\endcsname\relax
  \def\citenamefont#1{#1}\fi
\expandafter\ifx\csname url\endcsname\relax
  \def\url#1{\texttt{#1}}\fi
\expandafter\ifx\csname urlprefix\endcsname\relax\def\urlprefix{URL }\fi
\providecommand{\bibinfo}[2]{#2}
\providecommand{\eprint}[2][]{\url{#2}}

\bibitem[{\citenamefont{{Cameron}}(1959)}]{Cameron59ns}
\bibinfo{author}{\bibfnamefont{A.~G.} \bibnamefont{{Cameron}}},
  \bibinfo{journal}{\apj} \textbf{\bibinfo{volume}{130}}, \bibinfo{pages}{884}
  (\bibinfo{year}{1959}), \urlprefix\url{https://doi.org/10.1086/146780}.

\bibitem[{\citenamefont{Hewish et~al.}(1968)\citenamefont{Hewish, Bell,
  Pilkington, Scott, and Collins}}]{Hewish68oo}
\bibinfo{author}{\bibfnamefont{A.}~\bibnamefont{Hewish}},
  \bibinfo{author}{\bibfnamefont{S.~J.} \bibnamefont{Bell}},
  \bibinfo{author}{\bibfnamefont{J.~D.~H.} \bibnamefont{Pilkington}},
  \bibinfo{author}{\bibfnamefont{P.~F.} \bibnamefont{Scott}}, \bibnamefont{and}
  \bibinfo{author}{\bibfnamefont{R.~A.} \bibnamefont{Collins}},
  \bibinfo{journal}{Nature} \textbf{\bibinfo{volume}{217}}
  (\bibinfo{year}{1968}), \urlprefix\url{https://doi.org/10.1038/217709a0}.

\bibitem[{\citenamefont{Thorsett and Chakrabarty}(1999)}]{Thorsett99ns}
\bibinfo{author}{\bibfnamefont{S.~E.} \bibnamefont{Thorsett}} \bibnamefont{and}
  \bibinfo{author}{\bibfnamefont{D.}~\bibnamefont{Chakrabarty}},
  \bibinfo{journal}{Astrophys. J.} \textbf{\bibinfo{volume}{512}},
  \bibinfo{pages}{288} (\bibinfo{year}{1999}),
  \urlprefix\url{https://doi.org/10.1086/306742}.

\bibitem[{\citenamefont{{Barziv, O.} et~al.}(2001)\citenamefont{{Barziv, O.},
  {Kaper, L.}, {Van Kerkwijk, M. H.}, {Telting, J. H.}, and {Van Paradijs,
  J.}}}]{Barziv01tm}
\bibinfo{author}{\bibnamefont{{Barziv, O.}}},
  \bibinfo{author}{\bibnamefont{{Kaper, L.}}},
  \bibinfo{author}{\bibnamefont{{Van Kerkwijk, M. H.}}},
  \bibinfo{author}{\bibnamefont{{Telting, J. H.}}}, \bibnamefont{and}
  \bibinfo{author}{\bibnamefont{{Van Paradijs, J.}}}, \bibinfo{journal}{Astron.
  \& Astrophys.} \textbf{\bibinfo{volume}{377}}, \bibinfo{pages}{925}
  (\bibinfo{year}{2001}),
  \urlprefix\url{https://doi.org/10.1051/0004-6361:20011122}.

\bibitem[{\citenamefont{Demorest et~al.}(2010)\citenamefont{Demorest, Pennucci,
  Ransom, Roberts, and Hessels}}]{Demorest10}
\bibinfo{author}{\bibfnamefont{P.~B.} \bibnamefont{Demorest}},
  \bibinfo{author}{\bibfnamefont{T.}~\bibnamefont{Pennucci}},
  \bibinfo{author}{\bibfnamefont{S.~M.} \bibnamefont{Ransom}},
  \bibinfo{author}{\bibfnamefont{M.~S.~E.} \bibnamefont{Roberts}},
  \bibnamefont{and} \bibinfo{author}{\bibfnamefont{J.~W.~T.}
  \bibnamefont{Hessels}}, \bibinfo{journal}{Nature (London)}
  \textbf{\bibinfo{volume}{467}}, \bibinfo{pages}{1081} (\bibinfo{year}{2010}),
  \urlprefix\url{https://dx.doi.org/10.1038/nature09466}.

\bibitem[{\citenamefont{{van Kerkwijk} et~al.}(2011)\citenamefont{{van
  Kerkwijk}, {Breton}, and {Kulkarni}}}]{vanKerkwijk11ef}
\bibinfo{author}{\bibfnamefont{M.~H.} \bibnamefont{{van Kerkwijk}}},
  \bibinfo{author}{\bibfnamefont{R.~P.} \bibnamefont{{Breton}}},
  \bibnamefont{and} \bibinfo{author}{\bibfnamefont{S.~R.}
  \bibnamefont{{Kulkarni}}}, \bibinfo{journal}{\apj}
  \textbf{\bibinfo{volume}{728}}, \bibinfo{eid}{95} (\bibinfo{year}{2011}),
  \urlprefix\url{https://doi.org/10.1088/0004-637X/728/2/95}.

\bibitem[{\citenamefont{Antoniadis et~al.}(2013)\citenamefont{Antoniadis,
  Freire, Wex, Tauris, Lynch, van Kerkwijk, Kramer, Bassa, Dhillon, Driebe
  et~al.}}]{Antoniadis13}
\bibinfo{author}{\bibfnamefont{J.}~\bibnamefont{Antoniadis}},
  \bibinfo{author}{\bibfnamefont{P.~C.~C.} \bibnamefont{Freire}},
  \bibinfo{author}{\bibfnamefont{N.}~\bibnamefont{Wex}},
  \bibinfo{author}{\bibfnamefont{T.~M.} \bibnamefont{Tauris}},
  \bibinfo{author}{\bibfnamefont{R.~S.} \bibnamefont{Lynch}},
  \bibinfo{author}{\bibfnamefont{M.~H.} \bibnamefont{van Kerkwijk}},
  \bibinfo{author}{\bibfnamefont{M.}~\bibnamefont{Kramer}},
  \bibinfo{author}{\bibfnamefont{C.}~\bibnamefont{Bassa}},
  \bibinfo{author}{\bibfnamefont{V.~S.} \bibnamefont{Dhillon}},
  \bibinfo{author}{\bibfnamefont{T.}~\bibnamefont{Driebe}},
  \bibnamefont{et~al.}, \bibinfo{journal}{Science}
  \textbf{\bibinfo{volume}{340}}, \bibinfo{pages}{6131} (\bibinfo{year}{2013}),
  \urlprefix\url{https://doi.org/10.1126/science.1233232}.

\bibitem[{\citenamefont{{Cromartie} et~al.}(2020)\citenamefont{{Cromartie},
  {Fonseca}, {Ransom}, {Demorest}, {Arzoumanian}, {Blumer}, {Brook}, {DeCesar},
  {Dolch}, {Ellis} et~al.}}]{Cromartie20rs}
\bibinfo{author}{\bibfnamefont{H.~T.} \bibnamefont{{Cromartie}}},
  \bibinfo{author}{\bibfnamefont{E.}~\bibnamefont{{Fonseca}}},
  \bibinfo{author}{\bibfnamefont{S.~M.} \bibnamefont{{Ransom}}},
  \bibinfo{author}{\bibfnamefont{P.~B.} \bibnamefont{{Demorest}}},
  \bibinfo{author}{\bibfnamefont{Z.}~\bibnamefont{{Arzoumanian}}},
  \bibinfo{author}{\bibfnamefont{H.}~\bibnamefont{{Blumer}}},
  \bibinfo{author}{\bibfnamefont{P.~R.} \bibnamefont{{Brook}}},
  \bibinfo{author}{\bibfnamefont{M.~E.} \bibnamefont{{DeCesar}}},
  \bibinfo{author}{\bibfnamefont{T.}~\bibnamefont{{Dolch}}},
  \bibinfo{author}{\bibfnamefont{J.~A.} \bibnamefont{{Ellis}}},
  \bibnamefont{et~al.}, \bibinfo{journal}{Nature Astron.}
  \textbf{\bibinfo{volume}{4}}, \bibinfo{pages}{72} (\bibinfo{year}{2020}),
  \urlprefix\url{https://doi.org/10.1038/s41550-019-0880-2}.

\bibitem[{\citenamefont{Lattimer and Prakash}(2007)}]{Lattimer07}
\bibinfo{author}{\bibfnamefont{J.~M.} \bibnamefont{Lattimer}} \bibnamefont{and}
  \bibinfo{author}{\bibfnamefont{M.}~\bibnamefont{Prakash}},
  \bibinfo{journal}{Physics Reports} \textbf{\bibinfo{volume}{442}},
  \bibinfo{pages}{109 } (\bibinfo{year}{2007}), ISSN \bibinfo{issn}{0370-1573},
  \bibinfo{note}{the Hans Bethe Centennial Volume 1906-2006},
  \urlprefix\url{https://doi.org/10.1016/j.physrep.2007.02.003}.

\bibitem[{\citenamefont{{Heinke} et~al.}(2006)\citenamefont{{Heinke},
  {Rybicki}, {Narayan}, and {Grindlay}}}]{Heinke06}
\bibinfo{author}{\bibfnamefont{C.~O.} \bibnamefont{{Heinke}}},
  \bibinfo{author}{\bibfnamefont{G.~B.} \bibnamefont{{Rybicki}}},
  \bibinfo{author}{\bibfnamefont{R.}~\bibnamefont{{Narayan}}},
  \bibnamefont{and} \bibinfo{author}{\bibfnamefont{J.~E.}
  \bibnamefont{{Grindlay}}}, \bibinfo{journal}{Astrophys. J.}
  \textbf{\bibinfo{volume}{644}}, \bibinfo{pages}{1090} (\bibinfo{year}{2006}),
  \urlprefix\url{https://doi.org/10.1086/503701}.

\bibitem[{\citenamefont{Özel et~al.}(2009)\citenamefont{Özel, Güver, and
  Psaltis}}]{Ozel09}
\bibinfo{author}{\bibfnamefont{F.}~\bibnamefont{Özel}},
  \bibinfo{author}{\bibfnamefont{T.}~\bibnamefont{Güver}}, \bibnamefont{and}
  \bibinfo{author}{\bibfnamefont{D.}~\bibnamefont{Psaltis}},
  \bibinfo{journal}{Astrophys. J.} \textbf{\bibinfo{volume}{693}},
  \bibinfo{pages}{1775} (\bibinfo{year}{2009}),
  \urlprefix\url{https://doi.org/10.1088/0004-637x/693/2/1775}.

\bibitem[{\citenamefont{Pons et~al.}(2002)\citenamefont{Pons, Walter, Lattimer,
  Prakash, Neuhauser, and An}}]{Pons02}
\bibinfo{author}{\bibfnamefont{J.~A.} \bibnamefont{Pons}},
  \bibinfo{author}{\bibfnamefont{F.~M.} \bibnamefont{Walter}},
  \bibinfo{author}{\bibfnamefont{J.~M.} \bibnamefont{Lattimer}},
  \bibinfo{author}{\bibfnamefont{M.}~\bibnamefont{Prakash}},
  \bibinfo{author}{\bibfnamefont{R.}~\bibnamefont{Neuhauser}},
  \bibnamefont{and} \bibinfo{author}{\bibfnamefont{P.-h.} \bibnamefont{An}},
  \bibinfo{journal}{Astrophys. J.} \textbf{\bibinfo{volume}{564}},
  \bibinfo{pages}{981} (\bibinfo{year}{2002}),
  \urlprefix\url{https://doi.org/10.1086/324296}.

\bibitem[{\citenamefont{Ho and Heinke}(2009)}]{Ho09}
\bibinfo{author}{\bibfnamefont{W.~C.~G.} \bibnamefont{Ho}} \bibnamefont{and}
  \bibinfo{author}{\bibfnamefont{C.~O.} \bibnamefont{Heinke}},
  \bibinfo{journal}{Nature} \textbf{\bibinfo{volume}{462}}, \bibinfo{pages}{71}
  (\bibinfo{year}{2009}), \urlprefix\url{https://doi.org/0.1038/nature08525}.

\bibitem[{\citenamefont{Lattimer}(2012)}]{Lattimer12}
\bibinfo{author}{\bibfnamefont{J.~M.} \bibnamefont{Lattimer}},
  \bibinfo{journal}{Annu. Rev. Nucl. Part. Sci.} \textbf{\bibinfo{volume}{62}},
  \bibinfo{pages}{485} (\bibinfo{year}{2012}),
  \urlprefix\url{https://doi.org/10.1146/annurev-nucl-102711-095018}.

\bibitem[{\citenamefont{Ozel and Freire}(2016)}]{Ozel16}
\bibinfo{author}{\bibfnamefont{F.}~\bibnamefont{Ozel}} \bibnamefont{and}
  \bibinfo{author}{\bibfnamefont{P.}~\bibnamefont{Freire}},
  \bibinfo{journal}{Ann. Rev. Astron. Astrophys.}
  \textbf{\bibinfo{volume}{54}}, \bibinfo{pages}{401} (\bibinfo{year}{2016}),
  \urlprefix\url{https://doi.org/10.1146/annurev-astro-081915-023322}.

\bibitem[{\citenamefont{{LIGO Scientific Collab.} and {Virgo
  Collab.}}(2017)}]{Abbott17go}
\bibinfo{author}{\bibnamefont{{LIGO Scientific Collab.}}} \bibnamefont{and}
  \bibinfo{author}{\bibnamefont{{Virgo Collab.}}}, \bibinfo{journal}{Phys. Rev.
  Lett.} \textbf{\bibinfo{volume}{119}}, \bibinfo{pages}{161101}
  (\bibinfo{year}{2017}),
  \urlprefix\url{https://doi.org/10.1103/PhysRevLett.119.161101}.

\bibitem[{\citenamefont{{LIGO Scientific Collab.} and {Virgo
  Collab.}}(2018)}]{Abbott18mo}
\bibinfo{author}{\bibnamefont{{LIGO Scientific Collab.}}} \bibnamefont{and}
  \bibinfo{author}{\bibnamefont{{Virgo Collab.}}}, \bibinfo{journal}{Phys. Rev.
  Lett.} \textbf{\bibinfo{volume}{121}}, \bibinfo{pages}{161101}
  (\bibinfo{year}{2018}),
  \urlprefix\url{https://doi.org/10.1103/PhysRevLett.121.161101}.

\bibitem[{\citenamefont{Abbott et~al.}(2020{\natexlab{a}})}]{Abbott:2020uma}
\bibinfo{author}{\bibfnamefont{B.}~\bibnamefont{Abbott}} \bibnamefont{et~al.}
  (\bibinfo{collaboration}{LIGO Scientific, Virgo}),
  \bibinfo{journal}{Astrophys. J. Lett.} \textbf{\bibinfo{volume}{892}},
  \bibinfo{pages}{L3} (\bibinfo{year}{2020}{\natexlab{a}}),
  \urlprefix\url{https://doi.org/10.3847/2041-8213/ab75f5}.

\bibitem[{\citenamefont{Riley et~al.}(2019)\citenamefont{Riley, Watts,
  Bogdanov, Ray, Ludlam, Guillot, Arzoumanian, Baker, Bilous, Chakrabarty
  et~al.}}]{Riley2019}
\bibinfo{author}{\bibfnamefont{T.~E.} \bibnamefont{Riley}},
  \bibinfo{author}{\bibfnamefont{A.~L.} \bibnamefont{Watts}},
  \bibinfo{author}{\bibfnamefont{S.}~\bibnamefont{Bogdanov}},
  \bibinfo{author}{\bibfnamefont{P.~S.} \bibnamefont{Ray}},
  \bibinfo{author}{\bibfnamefont{R.~M.} \bibnamefont{Ludlam}},
  \bibinfo{author}{\bibfnamefont{S.}~\bibnamefont{Guillot}},
  \bibinfo{author}{\bibfnamefont{Z.}~\bibnamefont{Arzoumanian}},
  \bibinfo{author}{\bibfnamefont{C.~L.} \bibnamefont{Baker}},
  \bibinfo{author}{\bibfnamefont{A.~V.} \bibnamefont{Bilous}},
  \bibinfo{author}{\bibfnamefont{D.}~\bibnamefont{Chakrabarty}},
  \bibnamefont{et~al.}, \bibinfo{journal}{Astrophys. J.}
  \textbf{\bibinfo{volume}{887}}, \bibinfo{pages}{L21} (\bibinfo{year}{2019}),
  \urlprefix\url{https://doi.org/10.3847/2041-8213/ab481c}.

\bibitem[{\citenamefont{Miller et~al.}(2019)}]{Miller19}
\bibinfo{author}{\bibfnamefont{M.~C.} \bibnamefont{Miller}}
  \bibnamefont{et~al.}, \bibinfo{journal}{Astrophys. J.}
  \textbf{\bibinfo{volume}{887}}, \bibinfo{pages}{L24} (\bibinfo{year}{2019}),
  \urlprefix\url{https://doi.org/10.3847/2041-8213/ab50c5}.

\bibitem[{\citenamefont{Annala et~al.}(2018)\citenamefont{Annala, Gorda,
  Kurkela, and Vuorinen}}]{Annala18}
\bibinfo{author}{\bibfnamefont{E.}~\bibnamefont{Annala}},
  \bibinfo{author}{\bibfnamefont{T.}~\bibnamefont{Gorda}},
  \bibinfo{author}{\bibfnamefont{A.}~\bibnamefont{Kurkela}}, \bibnamefont{and}
  \bibinfo{author}{\bibfnamefont{A.}~\bibnamefont{Vuorinen}},
  \bibinfo{journal}{Phys. Rev. Lett.} \textbf{\bibinfo{volume}{120}},
  \bibinfo{pages}{172703} (\bibinfo{year}{2018}),
  \urlprefix\url{https://doi.org/10.1103/PhysRevLett.120.172703}.

\bibitem[{\citenamefont{Fattoyev et~al.}(2018)\citenamefont{Fattoyev,
  Piekarewicz, and Horowitz}}]{Fattoyev18}
\bibinfo{author}{\bibfnamefont{F.~J.} \bibnamefont{Fattoyev}},
  \bibinfo{author}{\bibfnamefont{J.}~\bibnamefont{Piekarewicz}},
  \bibnamefont{and} \bibinfo{author}{\bibfnamefont{C.~J.}
  \bibnamefont{Horowitz}}, \bibinfo{journal}{Phys. Rev. Lett.}
  \textbf{\bibinfo{volume}{120}}, \bibinfo{pages}{172702}
  (\bibinfo{year}{2018}),
  \urlprefix\url{https://doi.org/10.1103/PhysRevLett.120.172702}.

\bibitem[{\citenamefont{Kumar and Landry}(2019)}]{BKumar19}
\bibinfo{author}{\bibfnamefont{B.}~\bibnamefont{Kumar}} \bibnamefont{and}
  \bibinfo{author}{\bibfnamefont{P.}~\bibnamefont{Landry}},
  \bibinfo{journal}{Phys. Rev. D} \textbf{\bibinfo{volume}{99}},
  \bibinfo{pages}{123026} (\bibinfo{year}{2019}),
  \urlprefix\url{https://doi.org/10.1103/PhysRevD.99.123026}.

\bibitem[{\citenamefont{Coughlin et~al.}(2018)}]{Coughlin:2018miv}
\bibinfo{author}{\bibfnamefont{M.~W.} \bibnamefont{Coughlin}}
  \bibnamefont{et~al.}, \bibinfo{journal}{Mon. Not. Roy. Astron. Soc.}
  \textbf{\bibinfo{volume}{480}}, \bibinfo{pages}{3871} (\bibinfo{year}{2018}),
  \eprint{1805.09371}, \urlprefix\url{https://doi.org/10.1093/mnras/sty2174}.

\bibitem[{\citenamefont{Coughlin et~al.}(2019)\citenamefont{Coughlin, Dietrich,
  Margalit, and Metzger}}]{Coughlin:2018fis}
\bibinfo{author}{\bibfnamefont{M.~W.} \bibnamefont{Coughlin}},
  \bibinfo{author}{\bibfnamefont{T.}~\bibnamefont{Dietrich}},
  \bibinfo{author}{\bibfnamefont{B.}~\bibnamefont{Margalit}}, \bibnamefont{and}
  \bibinfo{author}{\bibfnamefont{B.~D.} \bibnamefont{Metzger}},
  \bibinfo{journal}{Mon. Not. Roy. Astron. Soc.}
  \textbf{\bibinfo{volume}{489}}, \bibinfo{pages}{L91} (\bibinfo{year}{2019}),
  \eprint{1812.04803},
  \urlprefix\url{https://dx.doi.org/10.1093/mnrasl/slz133}.

\bibitem[{\citenamefont{Tews et~al.}(2018)\citenamefont{Tews, Margueron, and
  Reddy}}]{Tews:2018iwm}
\bibinfo{author}{\bibfnamefont{I.}~\bibnamefont{Tews}},
  \bibinfo{author}{\bibfnamefont{J.}~\bibnamefont{Margueron}},
  \bibnamefont{and} \bibinfo{author}{\bibfnamefont{S.}~\bibnamefont{Reddy}},
  \bibinfo{journal}{Phys. Rev. C} \textbf{\bibinfo{volume}{98}},
  \bibinfo{pages}{045804} (\bibinfo{year}{2018}), \eprint{1804.02783},
  \urlprefix\url{https://doi.org/10.1103/PhysRevC.98.045804}.

\bibitem[{\citenamefont{Raithel}(2019)}]{Raithel19co}
\bibinfo{author}{\bibfnamefont{C.~A.} \bibnamefont{Raithel}},
  \bibinfo{journal}{Eur. Phys. J. A} \textbf{\bibinfo{volume}{55}},
  \bibinfo{pages}{80} (\bibinfo{year}{2019}),
  \urlprefix\url{https://doi.org/10.1140/epja/i2019-12759-5}.

\bibitem[{\citenamefont{De et~al.}(2018)\citenamefont{De, Finstad, Lattimer,
  Brown, Berger, and Biwer}}]{De18}
\bibinfo{author}{\bibfnamefont{S.}~\bibnamefont{De}},
  \bibinfo{author}{\bibfnamefont{D.}~\bibnamefont{Finstad}},
  \bibinfo{author}{\bibfnamefont{J.~M.} \bibnamefont{Lattimer}},
  \bibinfo{author}{\bibfnamefont{D.~A.} \bibnamefont{Brown}},
  \bibinfo{author}{\bibfnamefont{E.}~\bibnamefont{Berger}}, \bibnamefont{and}
  \bibinfo{author}{\bibfnamefont{C.~M.} \bibnamefont{Biwer}},
  \bibinfo{journal}{Phys. Rev. Lett.} \textbf{\bibinfo{volume}{121}},
  \bibinfo{pages}{091102} (\bibinfo{year}{2018}),
  \urlprefix\url{https://doi.org/10.1103/PhysRevLett.121.091102}.

\bibitem[{\citenamefont{Radice and Dai}(2019)}]{Radice2019}
\bibinfo{author}{\bibfnamefont{D.}~\bibnamefont{Radice}} \bibnamefont{and}
  \bibinfo{author}{\bibfnamefont{L.}~\bibnamefont{Dai}}, \bibinfo{journal}{Eur.
  Phys. J. A} \textbf{\bibinfo{volume}{55}} (\bibinfo{year}{2019}),
  \urlprefix\url{https://doi.org/10.1140/epja/i2019-12716-4}.

\bibitem[{\citenamefont{Capano et~al.}(2020)\citenamefont{Capano, Tews, Brown,
  Margalit, De, Kumar, Brown, Krishnan, and Reddy}}]{Capano20}
\bibinfo{author}{\bibfnamefont{C.~D.} \bibnamefont{Capano}},
  \bibinfo{author}{\bibfnamefont{I.}~\bibnamefont{Tews}},
  \bibinfo{author}{\bibfnamefont{S.~M.} \bibnamefont{Brown}},
  \bibinfo{author}{\bibfnamefont{B.}~\bibnamefont{Margalit}},
  \bibinfo{author}{\bibfnamefont{S.}~\bibnamefont{De}},
  \bibinfo{author}{\bibfnamefont{S.}~\bibnamefont{Kumar}},
  \bibinfo{author}{\bibfnamefont{D.~A.} \bibnamefont{Brown}},
  \bibinfo{author}{\bibfnamefont{B.}~\bibnamefont{Krishnan}}, \bibnamefont{and}
  \bibinfo{author}{\bibfnamefont{S.}~\bibnamefont{Reddy}},
  \bibinfo{journal}{Nature News}  (\bibinfo{year}{2020}),
  \urlprefix\url{https://doi.org/10.1038/s41550-020-1014-6}.

\bibitem[{\citenamefont{d'Etivaux et~al.}(2019)\citenamefont{d'Etivaux,
  Guillot, Margueron, Webb, Catelan, and Reisenegger}}]{Baillot2019}
\bibinfo{author}{\bibfnamefont{N.~B.} \bibnamefont{d'Etivaux}},
  \bibinfo{author}{\bibfnamefont{S.}~\bibnamefont{Guillot}},
  \bibinfo{author}{\bibfnamefont{J.}~\bibnamefont{Margueron}},
  \bibinfo{author}{\bibfnamefont{N.}~\bibnamefont{Webb}},
  \bibinfo{author}{\bibfnamefont{M.}~\bibnamefont{Catelan}}, \bibnamefont{and}
  \bibinfo{author}{\bibfnamefont{A.}~\bibnamefont{Reisenegger}},
  \bibinfo{journal}{Astrophys. J.} \textbf{\bibinfo{volume}{887}},
  \bibinfo{pages}{48} (\bibinfo{year}{2019}),
  \urlprefix\url{https://doi.org/10.3847/1538-4357/ab4f6c}.

\bibitem[{\citenamefont{Jiang et~al.}(2019)\citenamefont{Jiang, Tang, Shao,
  Han, Li, Wang, Jin, Fan, and Wei}}]{Jiang19}
\bibinfo{author}{\bibfnamefont{J.-L.} \bibnamefont{Jiang}},
  \bibinfo{author}{\bibfnamefont{S.-P.} \bibnamefont{Tang}},
  \bibinfo{author}{\bibfnamefont{D.-S.} \bibnamefont{Shao}},
  \bibinfo{author}{\bibfnamefont{M.-Z.} \bibnamefont{Han}},
  \bibinfo{author}{\bibfnamefont{Y.-J.} \bibnamefont{Li}},
  \bibinfo{author}{\bibfnamefont{Y.-Z.} \bibnamefont{Wang}},
  \bibinfo{author}{\bibfnamefont{Z.-P.} \bibnamefont{Jin}},
  \bibinfo{author}{\bibfnamefont{Y.-Z.} \bibnamefont{Fan}}, \bibnamefont{and}
  \bibinfo{author}{\bibfnamefont{D.-M.} \bibnamefont{Wei}},
  \bibinfo{journal}{The Astrophysical Journal} \textbf{\bibinfo{volume}{885}},
  \bibinfo{pages}{39} (\bibinfo{year}{2019}),
  \urlprefix\url{https://doi.org/10.3847/1538-4357/ab44b2}.

\bibitem[{\citenamefont{Dietrich et~al.}(2020)\citenamefont{Dietrich, Coughlin,
  Pang, Bulla, Heinzel, Issa, Tews, and Antier}}]{Dietrich:2020lps}
\bibinfo{author}{\bibfnamefont{T.}~\bibnamefont{Dietrich}},
  \bibinfo{author}{\bibfnamefont{M.~W.} \bibnamefont{Coughlin}},
  \bibinfo{author}{\bibfnamefont{P.~T.} \bibnamefont{Pang}},
  \bibinfo{author}{\bibfnamefont{M.}~\bibnamefont{Bulla}},
  \bibinfo{author}{\bibfnamefont{J.}~\bibnamefont{Heinzel}},
  \bibinfo{author}{\bibfnamefont{L.}~\bibnamefont{Issa}},
  \bibinfo{author}{\bibfnamefont{I.}~\bibnamefont{Tews}}, \bibnamefont{and}
  \bibinfo{author}{\bibfnamefont{S.}~\bibnamefont{Antier}}
  (\bibinfo{year}{2020}), \eprint{arXiv:2002.11355},
  \urlprefix\url{https://arxiv.org/abs/2002.11355}.

\bibitem[{\citenamefont{Landry and Essick}(2019)}]{Landry19}
\bibinfo{author}{\bibfnamefont{P.}~\bibnamefont{Landry}} \bibnamefont{and}
  \bibinfo{author}{\bibfnamefont{R.}~\bibnamefont{Essick}},
  \bibinfo{journal}{Phys. Rev. D} \textbf{\bibinfo{volume}{99}},
  \bibinfo{pages}{084049} (\bibinfo{year}{2019}),
  \urlprefix\url{https://doi.org/10.1103/PhysRevD.99.084049}.

\bibitem[{\citenamefont{Landry et~al.}(2020)\citenamefont{Landry, Essick, and
  Chatziioannou}}]{Landry20}
\bibinfo{author}{\bibfnamefont{P.}~\bibnamefont{Landry}},
  \bibinfo{author}{\bibfnamefont{R.}~\bibnamefont{Essick}}, \bibnamefont{and}
  \bibinfo{author}{\bibfnamefont{K.}~\bibnamefont{Chatziioannou}},
  \bibinfo{journal}{Phys. Rev. D} \textbf{\bibinfo{volume}{101}},
  \bibinfo{pages}{123007} (\bibinfo{year}{2020}),
  \urlprefix\url{https://doi.org/10.1103/PhysRevD.101.123007}.

\bibitem[{\citenamefont{Essick et~al.}(2020)\citenamefont{Essick, Tews, Landry,
  Reddy, and Holz}}]{Essick:2020flb}
\bibinfo{author}{\bibfnamefont{R.}~\bibnamefont{Essick}},
  \bibinfo{author}{\bibfnamefont{I.}~\bibnamefont{Tews}},
  \bibinfo{author}{\bibfnamefont{P.}~\bibnamefont{Landry}},
  \bibinfo{author}{\bibfnamefont{S.}~\bibnamefont{Reddy}}, \bibnamefont{and}
  \bibinfo{author}{\bibfnamefont{D.~E.} \bibnamefont{Holz}}
  (\bibinfo{year}{2020}), \eprint{arXiv:2004.07744},
  \urlprefix\url{https://arxiv.org/abs/2004.07744}.

\bibitem[{\citenamefont{Jiang et~al.}(2020)\citenamefont{Jiang, Tang, Wang,
  Fan, and Wei}}]{Jiang20}
\bibinfo{author}{\bibfnamefont{J.-L.} \bibnamefont{Jiang}},
  \bibinfo{author}{\bibfnamefont{S.-P.} \bibnamefont{Tang}},
  \bibinfo{author}{\bibfnamefont{Y.-Z.} \bibnamefont{Wang}},
  \bibinfo{author}{\bibfnamefont{Y.-Z.} \bibnamefont{Fan}}, \bibnamefont{and}
  \bibinfo{author}{\bibfnamefont{D.-M.} \bibnamefont{Wei}},
  \bibinfo{journal}{Astrophys. J.} \textbf{\bibinfo{volume}{892}},
  \bibinfo{pages}{55} (\bibinfo{year}{2020}),
  \urlprefix\url{https://doi.org/10.3847/1538-4357/ab77cf}.

\bibitem[{\citenamefont{Lattimer}(2019)}]{Lattimer2019}
\bibinfo{author}{\bibfnamefont{J.~M.} \bibnamefont{Lattimer}},
  \bibinfo{journal}{Universe} \textbf{\bibinfo{volume}{5}},
  \bibinfo{pages}{159} (\bibinfo{year}{2019}), ISSN \bibinfo{issn}{2218-1997},
  \urlprefix\url{http://dx.doi.org/10.3390/universe5070159}.

\bibitem[{\citenamefont{Ayriyan et~al.}(2019)\citenamefont{Ayriyan,
  Alvarez-Castillo, Blaschke, and Grigorian}}]{Ayrian18}
\bibinfo{author}{\bibfnamefont{A.}~\bibnamefont{Ayriyan}},
  \bibinfo{author}{\bibfnamefont{D.}~\bibnamefont{Alvarez-Castillo}},
  \bibinfo{author}{\bibfnamefont{D.}~\bibnamefont{Blaschke}}, \bibnamefont{and}
  \bibinfo{author}{\bibfnamefont{H.}~\bibnamefont{Grigorian}},
  \bibinfo{journal}{Universe} \textbf{\bibinfo{volume}{5}}
  (\bibinfo{year}{2019}), ISSN \bibinfo{issn}{2218-1997},
  \urlprefix\url{https://doi.org/10.3390/universe5020061}.

\bibitem[{\citenamefont{Raaijmakers et~al.}(2020)}]{Raaijmakers:2019dks}
\bibinfo{author}{\bibfnamefont{G.}~\bibnamefont{Raaijmakers}}
  \bibnamefont{et~al.}, \bibinfo{journal}{Astrophys. J. Lett.}
  \textbf{\bibinfo{volume}{893}}, \bibinfo{pages}{L21} (\bibinfo{year}{2020}),
  \urlprefix\url{https://doi.org/10.3847/2041-8213/ab822f}.

\bibitem[{\citenamefont{Raithel et~al.}(2020)\citenamefont{Raithel, Ozel, and
  Psaltis}}]{Raithel20os}
\bibinfo{author}{\bibfnamefont{C.}~\bibnamefont{Raithel}},
  \bibinfo{author}{\bibfnamefont{F.}~\bibnamefont{Ozel}}, \bibnamefont{and}
  \bibinfo{author}{\bibfnamefont{D.}~\bibnamefont{Psaltis}}
  (\bibinfo{year}{2020}), \eprint{arXiv:2004.00656},
  \urlprefix\url{https://arxiv.org/abs/2004.00656}.

\bibitem[{\citenamefont{Steiner et~al.}(2015)\citenamefont{Steiner, Gandolfi,
  Fattoyev, and Newton}}]{Steiner15un}
\bibinfo{author}{\bibfnamefont{A.~W.} \bibnamefont{Steiner}},
  \bibinfo{author}{\bibfnamefont{S.}~\bibnamefont{Gandolfi}},
  \bibinfo{author}{\bibfnamefont{F.~J.} \bibnamefont{Fattoyev}},
  \bibnamefont{and} \bibinfo{author}{\bibfnamefont{W.~G.}
  \bibnamefont{Newton}}, \bibinfo{journal}{Phys. Rev. C}
  \textbf{\bibinfo{volume}{91}}, \bibinfo{pages}{015804}
  (\bibinfo{year}{2015}),
  \urlprefix\url{https://doi.org/10.1103/PhysRevC.91.015804}.

\bibitem[{\citenamefont{Steiner et~al.}(2010)\citenamefont{Steiner, Lattimer,
  and {Brown}}}]{Steiner10te}
\bibinfo{author}{\bibfnamefont{A.~W.} \bibnamefont{Steiner}},
  \bibinfo{author}{\bibfnamefont{J.~M.} \bibnamefont{Lattimer}},
  \bibnamefont{and} \bibinfo{author}{\bibfnamefont{E.~F.}
  \bibnamefont{{Brown}}}, \bibinfo{journal}{Astrophys. J.}
  \textbf{\bibinfo{volume}{722}}, \bibinfo{pages}{33} (\bibinfo{year}{2010}),
  \urlprefix\url{https://doi.org/10.1088/0004-637X/722/1/33}.

\bibitem[{\citenamefont{Steiner and Gandolfi}(2012)}]{Steiner12cn}
\bibinfo{author}{\bibfnamefont{A.~W.} \bibnamefont{Steiner}} \bibnamefont{and}
  \bibinfo{author}{\bibfnamefont{S.}~\bibnamefont{Gandolfi}},
  \bibinfo{journal}{Phys. Rev. Lett.} \textbf{\bibinfo{volume}{108}},
  \bibinfo{pages}{081102} (\bibinfo{year}{2012}),
  \urlprefix\url{https://doi.org/10.1103/PhysRevLett.108.081102}.

\bibitem[{\citenamefont{{Read} et~al.}(2009)\citenamefont{{Read}, {Lackey},
  {Owen}, and {Friedman}}}]{Read09co}
\bibinfo{author}{\bibfnamefont{J.~S.} \bibnamefont{{Read}}},
  \bibinfo{author}{\bibfnamefont{B.~D.} \bibnamefont{{Lackey}}},
  \bibinfo{author}{\bibfnamefont{B.~J.} \bibnamefont{{Owen}}},
  \bibnamefont{and} \bibinfo{author}{\bibfnamefont{J.~L.}
  \bibnamefont{{Friedman}}}, \bibinfo{journal}{\prd}
  \textbf{\bibinfo{volume}{79}}, \bibinfo{eid}{124032} (\bibinfo{year}{2009}),
  \eprint{0812.2163},
  \urlprefix\url{https://doi.org/10.1103/PhysRevD.79.124032}.

\bibitem[{\citenamefont{Lattimer and Steiner}(2014)}]{Lattimer14co}
\bibinfo{author}{\bibfnamefont{J.~M.} \bibnamefont{Lattimer}} \bibnamefont{and}
  \bibinfo{author}{\bibfnamefont{A.~W.} \bibnamefont{Steiner}},
  \bibinfo{journal}{Eur. Phys. J. A} \textbf{\bibinfo{volume}{50}},
  \bibinfo{pages}{40} (\bibinfo{year}{2014}),
  \urlprefix\url{https://doi.org/10.1140/epja/i2014-14040-y}.

\bibitem[{\citenamefont{Steiner et~al.}(2013)\citenamefont{Steiner, Lattimer,
  and Brown}}]{Steiner13tn}
\bibinfo{author}{\bibfnamefont{A.~W.} \bibnamefont{Steiner}},
  \bibinfo{author}{\bibfnamefont{J.~M.} \bibnamefont{Lattimer}},
  \bibnamefont{and} \bibinfo{author}{\bibfnamefont{E.~F.} \bibnamefont{Brown}},
  \bibinfo{journal}{Astrophys. J. Lett.} \textbf{\bibinfo{volume}{765}},
  \bibinfo{pages}{5} (\bibinfo{year}{2013}),
  \urlprefix\url{https://doi.org/10.1088/2041-8205/765/1/L5}.

\bibitem[{\citenamefont{Greif et~al.}(2019)\citenamefont{Greif, Raaijmakers,
  Hebeler, Schwenk, and Watts}}]{Greif:2018njt}
\bibinfo{author}{\bibfnamefont{S.}~\bibnamefont{Greif}},
  \bibinfo{author}{\bibfnamefont{G.}~\bibnamefont{Raaijmakers}},
  \bibinfo{author}{\bibfnamefont{K.}~\bibnamefont{Hebeler}},
  \bibinfo{author}{\bibfnamefont{A.}~\bibnamefont{Schwenk}}, \bibnamefont{and}
  \bibinfo{author}{\bibfnamefont{A.}~\bibnamefont{Watts}},
  \bibinfo{journal}{Mon. Not. Roy. Astron. Soc.}
  \textbf{\bibinfo{volume}{485}}, \bibinfo{pages}{5363} (\bibinfo{year}{2019}),
  \eprint{1812.08188}, \urlprefix\url{https://doi.org/10.1093/mnras/stz654}.

\bibitem[{\citenamefont{Steiner et~al.}(2018)\citenamefont{Steiner, Heinke,
  Bogdanov, Li, Ho, Bahramian, and Han}}]{Steiner18ct}
\bibinfo{author}{\bibfnamefont{A.~W.} \bibnamefont{Steiner}},
  \bibinfo{author}{\bibfnamefont{C.~O.} \bibnamefont{Heinke}},
  \bibinfo{author}{\bibfnamefont{S.}~\bibnamefont{Bogdanov}},
  \bibinfo{author}{\bibfnamefont{C.}~\bibnamefont{Li}},
  \bibinfo{author}{\bibfnamefont{W.~C.~G.} \bibnamefont{Ho}},
  \bibinfo{author}{\bibfnamefont{A.}~\bibnamefont{Bahramian}},
  \bibnamefont{and} \bibinfo{author}{\bibfnamefont{S.}~\bibnamefont{Han}},
  \bibinfo{journal}{Mon. Not. Roy. Astron. Soc.}
  \textbf{\bibinfo{volume}{476}}, \bibinfo{pages}{421} (\bibinfo{year}{2018}),
  \urlprefix\url{https://doi.org/10.1093/mnras/sty215}.

\bibitem[{\citenamefont{Tolman}(1939)}]{Tolman:1939jz}
\bibinfo{author}{\bibfnamefont{R.~C.} \bibnamefont{Tolman}},
  \bibinfo{journal}{Phys. Rev.} \textbf{\bibinfo{volume}{55}},
  \bibinfo{pages}{364} (\bibinfo{year}{1939}),
  \urlprefix\url{https://doi.org/10.1103/PhysRev.55.364}.

\bibitem[{\citenamefont{Oppenheimer and Volkoff}(1939)}]{Oppenheimer:1939ne}
\bibinfo{author}{\bibfnamefont{J.}~\bibnamefont{Oppenheimer}} \bibnamefont{and}
  \bibinfo{author}{\bibfnamefont{G.}~\bibnamefont{Volkoff}},
  \bibinfo{journal}{Phys. Rev.} \textbf{\bibinfo{volume}{55}},
  \bibinfo{pages}{374} (\bibinfo{year}{1939}),
  \urlprefix\url{https://doi.org/10.1103/PhysRev.55.374}.

\bibitem[{\citenamefont{Yagi and Yunes}(2013)}]{Yagi13}
\bibinfo{author}{\bibfnamefont{K.}~\bibnamefont{Yagi}} \bibnamefont{and}
  \bibinfo{author}{\bibfnamefont{N.}~\bibnamefont{Yunes}},
  \bibinfo{journal}{Science} \textbf{\bibinfo{volume}{341}},
  \bibinfo{pages}{365} (\bibinfo{year}{2013}),
  \urlprefix\url{https://doi.org/10.1126/science.1236462}.

\bibitem[{\citenamefont{Steiner et~al.}(2016)\citenamefont{Steiner, Lattimer,
  and Brown}}]{Steiner16ns}
\bibinfo{author}{\bibfnamefont{A.~W.} \bibnamefont{Steiner}},
  \bibinfo{author}{\bibfnamefont{J.~M.} \bibnamefont{Lattimer}},
  \bibnamefont{and} \bibinfo{author}{\bibfnamefont{E.~F.} \bibnamefont{Brown}},
  \bibinfo{journal}{Eur. Phys. J. A} \textbf{\bibinfo{volume}{52}},
  \bibinfo{pages}{18} (\bibinfo{year}{2016}),
  \urlprefix\url{https://doi.org/10.1140/epja/i2016-16018-1}.

\bibitem[{\citenamefont{Han and Steiner}(2019)}]{Han19td}
\bibinfo{author}{\bibfnamefont{S.}~\bibnamefont{Han}} \bibnamefont{and}
  \bibinfo{author}{\bibfnamefont{A.~W.} \bibnamefont{Steiner}},
  \bibinfo{journal}{Phys. Rev. D} \textbf{\bibinfo{volume}{99}},
  \bibinfo{pages}{083014} (\bibinfo{year}{2019}),
  \urlprefix\url{https://doi.org/10.1103/PhysRevD.99.083014}.

\bibitem[{\citenamefont{Carson et~al.}(2019)\citenamefont{Carson,
  Chatziioannou, Haster, Yagi, and Yunes}}]{Carson:2019rjx}
\bibinfo{author}{\bibfnamefont{Z.}~\bibnamefont{Carson}},
  \bibinfo{author}{\bibfnamefont{K.}~\bibnamefont{Chatziioannou}},
  \bibinfo{author}{\bibfnamefont{C.-J.} \bibnamefont{Haster}},
  \bibinfo{author}{\bibfnamefont{K.}~\bibnamefont{Yagi}}, \bibnamefont{and}
  \bibinfo{author}{\bibfnamefont{N.}~\bibnamefont{Yunes}},
  \bibinfo{journal}{Phys. Rev. D} \textbf{\bibinfo{volume}{99}},
  \bibinfo{pages}{083016} (\bibinfo{year}{2019}),
  \urlprefix\url{https://doi.org/10.1103/PhysRevD.99.083016}.

\bibitem[{\citenamefont{N\"{a}ttil\"{a}
  et~al.}(2016)\citenamefont{N\"{a}ttil\"{a}, Steiner, Kajava, Suleimanov, and
  Poutanen}}]{Nattila16eo}
\bibinfo{author}{\bibfnamefont{J.}~\bibnamefont{N\"{a}ttil\"{a}}},
  \bibinfo{author}{\bibfnamefont{A.~W.} \bibnamefont{Steiner}},
  \bibinfo{author}{\bibfnamefont{J.~J.~E.} \bibnamefont{Kajava}},
  \bibinfo{author}{\bibfnamefont{V.~F.} \bibnamefont{Suleimanov}},
  \bibnamefont{and} \bibinfo{author}{\bibfnamefont{J.}~\bibnamefont{Poutanen}},
  \bibinfo{journal}{Astron. Astrophys.} \textbf{\bibinfo{volume}{591}},
  \bibinfo{pages}{A25} (\bibinfo{year}{2016}),
  \urlprefix\url{https://doi.org/10.1051/0004-6361/201527416}.

\bibitem[{\citenamefont{N\"{a}ttil\"{a}
  et~al.}(2017)\citenamefont{N\"{a}ttil\"{a}, Miller, Steiner, Kajava,
  Suleimanov, and Poutanen}}]{Nattila17ns}
\bibinfo{author}{\bibfnamefont{J.}~\bibnamefont{N\"{a}ttil\"{a}}},
  \bibinfo{author}{\bibfnamefont{M.~C.} \bibnamefont{Miller}},
  \bibinfo{author}{\bibfnamefont{A.~W.} \bibnamefont{Steiner}},
  \bibinfo{author}{\bibfnamefont{J.~J.~E.} \bibnamefont{Kajava}},
  \bibinfo{author}{\bibfnamefont{V.~F.} \bibnamefont{Suleimanov}},
  \bibnamefont{and} \bibinfo{author}{\bibfnamefont{J.}~\bibnamefont{Poutanen}},
  \bibinfo{journal}{Astron. and Astrophys.} \textbf{\bibinfo{volume}{608}},
  \bibinfo{pages}{A31} (\bibinfo{year}{2017}),
  \urlprefix\url{https:/doi.org/10.1051/0004-6361/201731082}.

\bibitem[{\citenamefont{{LIGO Scientific Collab.} and {Virgo
  Collab.}}(2019)}]{Abbott19}
\bibinfo{author}{\bibnamefont{{LIGO Scientific Collab.}}} \bibnamefont{and}
  \bibinfo{author}{\bibnamefont{{Virgo Collab.}}}, \bibinfo{journal}{Phys. Rev.
  X} \textbf{\bibinfo{volume}{9}}, \bibinfo{pages}{011001}
  (\bibinfo{year}{2019}),
  \urlprefix\url{https://doi.org/10.1103/PhysRevX.9.011001}.

\bibitem[{\citenamefont{{Lange} et~al.}(2018)\citenamefont{{Lange},
  {O'Shaughnessy}, and {Rizzo}}}]{RIFT}
\bibinfo{author}{\bibfnamefont{J.}~\bibnamefont{{Lange}}},
  \bibinfo{author}{\bibfnamefont{R.}~\bibnamefont{{O'Shaughnessy}}},
  \bibnamefont{and} \bibinfo{author}{\bibfnamefont{M.}~\bibnamefont{{Rizzo}}},
  \bibinfo{journal}{arXiv:1805.10457}  (\bibinfo{year}{2018}),
  \urlprefix\url{https://arxiv.org/abs/1805.10457}.

\bibitem[{\citenamefont{Abbott et~al.}(2020{\natexlab{b}})}]{Abbott:2020khf}
\bibinfo{author}{\bibfnamefont{R.}~\bibnamefont{Abbott}} \bibnamefont{et~al.}
  (\bibinfo{collaboration}{LIGO Scientific, Virgo}),
  \bibinfo{journal}{Astrophys. J. Lett.} \textbf{\bibinfo{volume}{896}},
  \bibinfo{pages}{L44} (\bibinfo{year}{2020}{\natexlab{b}}),
  \urlprefix\url{https://doi.org/10.3847/2041-8213/ab960f}.

\bibitem[{\citenamefont{Essick and Landry}(2020)}]{Essick:2020ghc}
\bibinfo{author}{\bibfnamefont{R.}~\bibnamefont{Essick}} \bibnamefont{and}
  \bibinfo{author}{\bibfnamefont{P.}~\bibnamefont{Landry}}
  (\bibinfo{year}{2020}), \eprint{arXiv:2007.01372},
  \urlprefix\url{https://arxiv.org/abs/2007.01372}.

\bibitem[{\citenamefont{Tews et~al.}(2020)\citenamefont{Tews, Pang, Dietrich,
  Coughlin, Antier, Bulla, Heinzel, and Issa}}]{Tews:2020ylw}
\bibinfo{author}{\bibfnamefont{I.}~\bibnamefont{Tews}},
  \bibinfo{author}{\bibfnamefont{P.~T.} \bibnamefont{Pang}},
  \bibinfo{author}{\bibfnamefont{T.}~\bibnamefont{Dietrich}},
  \bibinfo{author}{\bibfnamefont{M.~W.} \bibnamefont{Coughlin}},
  \bibinfo{author}{\bibfnamefont{S.}~\bibnamefont{Antier}},
  \bibinfo{author}{\bibfnamefont{M.}~\bibnamefont{Bulla}},
  \bibinfo{author}{\bibfnamefont{J.}~\bibnamefont{Heinzel}}, \bibnamefont{and}
  \bibinfo{author}{\bibfnamefont{L.}~\bibnamefont{Issa}}
  (\bibinfo{year}{2020}), \eprint{arXiv:2007.06057},
  \urlprefix\url{https://arxiv.org/abs/2007.06057}.

\bibitem[{\citenamefont{Radice et~al.}(2018)\citenamefont{Radice, Perego,
  Zappa, and Bernuzzi}}]{Radice17gj}
\bibinfo{author}{\bibfnamefont{D.}~\bibnamefont{Radice}},
  \bibinfo{author}{\bibfnamefont{A.}~\bibnamefont{Perego}},
  \bibinfo{author}{\bibfnamefont{F.}~\bibnamefont{Zappa}}, \bibnamefont{and}
  \bibinfo{author}{\bibfnamefont{S.}~\bibnamefont{Bernuzzi}},
  \bibinfo{journal}{Astrophys. J. Lett.} \textbf{\bibinfo{volume}{852}},
  \bibinfo{pages}{L29} (\bibinfo{year}{2018}),
  \urlprefix\url{https://doi.org/10.3847/2041-8213/aaa402}.

\bibitem[{\citenamefont{Bauswein and Stergioulas}(2017)}]{Bauswein17sd}
\bibinfo{author}{\bibfnamefont{A.}~\bibnamefont{Bauswein}} \bibnamefont{and}
  \bibinfo{author}{\bibfnamefont{N.}~\bibnamefont{Stergioulas}},
  \bibinfo{journal}{Mon. Not. Roy. Astron. Soc.}
  \textbf{\bibinfo{volume}{471}}, \bibinfo{pages}{4956} (\bibinfo{year}{2017}),
  \eprint{1702.02567}, \urlprefix\url{https://doi.org/10.1093/mnras/stx1983}.

\bibitem[{\citenamefont{Kiuchi et~al.}(2019)\citenamefont{Kiuchi, Kyutoku,
  Shibata, and Taniguchi}}]{Kiuchi19rt}
\bibinfo{author}{\bibfnamefont{K.}~\bibnamefont{Kiuchi}},
  \bibinfo{author}{\bibfnamefont{K.}~\bibnamefont{Kyutoku}},
  \bibinfo{author}{\bibfnamefont{M.}~\bibnamefont{Shibata}}, \bibnamefont{and}
  \bibinfo{author}{\bibfnamefont{K.}~\bibnamefont{Taniguchi}},
  \bibinfo{journal}{Astrophys. J. Lett.} \textbf{\bibinfo{volume}{876}},
  \bibinfo{pages}{L31} (\bibinfo{year}{2019}), \eprint{1903.01466},
  \urlprefix\url{https://doi.org/10.3847/2041-8213/ab1e45}.

\bibitem[{\citenamefont{Ruiz et~al.}(2018)\citenamefont{Ruiz, Shapiro, and
  Tsokaros}}]{Ruiz18}
\bibinfo{author}{\bibfnamefont{M.}~\bibnamefont{Ruiz}},
  \bibinfo{author}{\bibfnamefont{S.~L.} \bibnamefont{Shapiro}},
  \bibnamefont{and} \bibinfo{author}{\bibfnamefont{A.}~\bibnamefont{Tsokaros}},
  \bibinfo{journal}{Phys. Rev. D} \textbf{\bibinfo{volume}{97}},
  \bibinfo{pages}{021501} (\bibinfo{year}{2018}),
  \urlprefix\url{https://doi.org/10.1103/PhysRevD.97.021501}.

\bibitem[{\citenamefont{Steiner}(2014{\natexlab{a}})}]{Steiner14ba}
\bibinfo{author}{\bibfnamefont{A.~W.} \bibnamefont{Steiner}},
  \emph{\bibinfo{title}{bamr: Bayesian analysis of mass and radius
  observations}} (\bibinfo{year}{2014}{\natexlab{a}}),
  \bibinfo{note}{{Astrophysics Source Code Library}, record ascl:1408.020},
  \urlprefix\url{http://ascl.net/1408.020}.

\bibitem[{\citenamefont{Steiner}(2014{\natexlab{b}})}]{Steiner14oo}
\bibinfo{author}{\bibfnamefont{A.~W.} \bibnamefont{Steiner}},
  \emph{\bibinfo{title}{O2scl: Object-oriented scientific computing library}}
  (\bibinfo{year}{2014}{\natexlab{b}}), \bibinfo{note}{{Astrophysics Source
  Code Library}, record ascl:1408.019},
  \urlprefix\url{http://ascl.net/1408.019}.

\bibitem[{\citenamefont{Frigo and Johnson}(2012)}]{Frigo12}
\bibinfo{author}{\bibfnamefont{F.}~\bibnamefont{Frigo}} \bibnamefont{and}
  \bibinfo{author}{\bibfnamefont{S.~G.} \bibnamefont{Johnson}},
  \emph{\bibinfo{title}{Fftw: Fastest fourier transform in the west}}
  (\bibinfo{year}{2012}), \bibinfo{note}{{FFTW: Fastest Fourier Transform in
  the West}, record ascl:1201.015}, \urlprefix\url{http://ascl.net/1201.015}.

\end{thebibliography}

\end{document}